\documentclass[12pt]{article}
\usepackage{amsmath,amssymb,setspace,geometry,verbatim,graphicx}
\usepackage{url}
\usepackage{fancybox}

\begin{document}

\centerline{\Large\bf Statistical Methods for Astronomy}

\bigskip
\centerline{\large\bf Eric D. Feigelson$^{1,3}$ and G. Jogesh Babu$^{2,3}$}

\bigskip\noindent 
$^1$ Department of Astronomy \& Astrophysics, Pennsylvania State University, University Park PA 16802 USA \\
$^2$ Department of Statistics, Pennsylvania State University, University Park PA, 16802, USA \\
$^3$ Center for Astrostatistics, Pennsylvania State University, University Park PA 16802, USA

\bigskip

\begin{abstract}
This review outlines concepts of mathematical statistics,
elements of probability theory, hypothesis tests and point estimation for use in the analysis
of modern astronomical data.  Least squares, maximum 
likelihood, and Bayesian approaches to statistical inference are treated.  Resampling 
methods, particularly the bootstrap, provide valuable 
procedures when distributions functions of statistics are not known. Several
approaches to model selection and goodness of fit are considered.

Applied statistics relevant to astronomical research are briefly
discussed:  nonparametric methods for use when
little is known about the behavior of the astronomical populations or processes; 
data smoothing with kernel density estimation and nonparametric regression;
unsupervised clustering and supervised classification procedures for multivariate
problems; survival analysis for astronomical datasets with nondetections; 
time- and frequency-domain times series analysis for light curves; and
spatial statistics to interpret the spatial distributions of points in low dimensions.
Two types of resources are presented: about 40 recommended texts and monographs 
in various fields of statistics, and   Second, the public domain {\bf R} software system for statistical analysis.
Together with its $\sim 3500$ (and growing) add-on CRAN packages, {\bf R} implements a vast range of
statistical procedures in a coherent high-level language with advanced graphics. 
\end{abstract}

\section{Role and history of statistics in astronomy}

Through much of the $20^{th}$ century, astronomers generally viewed statistical me\-thod\-ology as an established collection of mechanical tools to assist in the analysis of quantitative data.  A narrow suite of classical methods were commonly used, such as model fitting by minimizing a $\chi^2$-like statistic, goodness of fit tests of a model to a dataset with the Kolmogorov-Smirnov statistic, and Fourier analysis of time series.  These methods are often used beyond their range of applicability.  Except for a vanguard of astronomers expert in specific advanced techniques (e.g., Bayesian or wavelet analysis), there was little awareness that statistical methodology had progressed considerably in recent decades to provide a wealth of techniques for wide range of problems.   Interest in astrostatistics and associated fields like astroinformatics is rapidly growing today.  

The role of statistical analysis as an element of scientific inference has been widely debated (Rao 1997). 
Some statisticians feel that, while statistical characterization can be effective, statistical modeling is often unreliable.  Statistician G. E. P. Box famously said `Essentially, all models are wrong, but some are useful', and Sir D. R. Cox (2006) wrote `The use, if any, in the process of simple {\it quantitative} notions of probability and their numerical assessment [to scientific inference] is unclear'.  Others are more optimistic, astrostatistician P. C. Gregory (2005) writes:

\begin{quote} Our [scientific] understanding comes through the development of theoretical models which are capable of explaining the existing observations as well as making testable predictions.  ... Fortunately, a variety of sophisticated mathematical and computational approaches have been developed to help us through this interface, these go under the general heading of statistical inference.  \end{quote}

Astronomers might distinguish cases where the model has a strong astrophysical underpinning (such as fitting a Keplerian ellipse to a planetary orbit) and cases where the model does not have a clear astrophysical explanation (such as fitting a power law to the distribution of stellar masses).  In all cases, astronomers should carefully formulate the question to be addressed, apply carefully chosen statistical approaches to the dataset to address this question with clearly stated assumptions, and recognize that the link between the statistical findings and reality may not be straightforward.  

Prior to the $20^{th}$ century, many statistical developments were centered around astronomical problems (Stigler 1986).  Ancient Greek, Arabic, and Renaissance astronomers debated how to estimate a quantity, such as the length of a solar year, based on repeated and inconsistent measurements.  Most favored the middle of the extreme values, and some scholars feared that repeated measurement led to greater, not reduced, uncertainty.  The mean was promoted by Tycho Brahe and Galileo Galilei, but did not become the standard procedure until the mid-18$^{th}$ century.  The $11^{th}$ century Persian astronomer al-Biruni and Galileo discussed propagation of measurement errors.  Adrian Legendre and Carl Friedrich Gauss developed the `Gaussian' or normal error distribution in the early $19^{th}$ century to address discrepant measurement in celestial mechanics.  The normal distribution was intertwined with the least-squares estimation technique developed by Legendre and Pierre-Simon Laplace.  Leading astronomers throughout Europe contributed to least-squares theory during the $19^{th}$ century.  

However, the close association of statistics with astronomy atrophied during the beginning of the $20^{th}$ century.  During the middle of the century, astronomers continued using least-squares techniques, although heuristic procedures were also common.  Astronomers did not adopt the powerful methods of maximum likelihood estimation (MLE), formulated by Sir R. A. Fisher in the 1920s and widely promulgated in other fields by the 1950s.  MLE came into use during the 1970-80s, along with the nonparametric Kolmogorov-Smirnov statistic.  Practical volumes by Bevington (1969) and Press et al. (1986) with Fortran source codes promulgated a suite of classical methods.   The adoption of a wider range of statistical methodology began during the 1990s with the emergence of cross-disciplinary interactions between astronomers and statisticians including collaborative research groups, conferences, didactic summer schools and software resources. A small but growing research field of astrostatistics was estalbished.   The activity is propelled both by the sophistication of data analysis and modeling problems, and the exponentially growing quantity of publicly available astronomical data.  

The purpose of this review is to introduce some of the concepts and results of a broad scope of modern statistical methodology that can be effective for astronomical data and science analysis.  Section 2 outlines concepts and results of statistical inference that underlie statistical analysis of astronomical data.  Section 3 reviews several fields of applied statistics relevant for astronomy.  Section 4 discusses resources available for the astronomer to learn appropriate statistical methodology, including the powerful {\bf R} software system.  Key terminology is noted in quotation marks (e.g., `bootstrap resampling').  The coverage is abbreviated and not complete in any fashion.  Topics are mostly restricted to areas of established statistical methodology relevant to astronomy, recent advances in the astronomical literature are mostly not considered.  A more comprehensive treatment for astronomers can be found in Feigelson \& Babu (2011), and the volumes listed in the Table~\ref{statbooks.tbl} should be consulted prior to serious statistical investigations.

\section{Statistical inference}

\subsection{Concepts of statistical inference}

A random variable' is a function of potential outcomes of an experiment or population. A particular realization of a random variable is often called a `data point'.  The magnitude of stars or redshifts of galaxies are examples of random variables.  An astronomical dataset might  contain photometric measurements of a sample of stars, spectroscopic redshifts of a sample of galaxies, categorical measurements (such as radio-loud and radio-quiet active galactic nuclei), or brightness measurements as a function of sky location (an image), wavelength of light (a spectrum), or of time (a light curve).   The dataset might be very small so that large-$N$ approximations do not apply, or very large making computations difficult to perform.  Observed values may be accompanied by secondary information, such as estimates of the errors arising from the measurement process.  `Statistics' are functions of random variables, ranging from simple function, such as the mean value, to complicated functions, such as an adaptive kernel smoother with cross-validation bandwidths and bootstrap errors (Figure~\ref{PSSS_Feigelson_adapt.fig} below).  

Statisticians have established the distributional properties of a number of statistics through formal mathematical theorems.  For example, under broad conditions, the `Central Limit Theorem' indicates that the mean value of a sufficiently large sample of independent random variables is normally distributed.   Astronomers can invent statistics that reveal some scientifically interesting properties of a dataset, but can not assume they follow simple distributions unless this has been established by theorems.  But in many cases, Monte Carlo methods such as the bootstrap can numerically recover the distribution of the statistic from a particular dataset under study.  

Statistical inference, in principle, helps reach conclusions that extend beyond the immediate data to derive broadly applicable insights into the underlying population.  The field of statistical inference is very large and can be classified in a number of ways.  `Nonparametric inference' gives probabilistic statements about the data which do not assume any particular distribution (e.g., Gaussian, power law) or parametric model for the data, while `parametric inference' assumes some distributions or functional relationships.  These relationships can be simple heuristic relations, as in linear regression, or can be complex functions derived from astrophysical theory.   Inference can be viewed as the combination of two basic branches: `point estimation' (such as estimating the mean of a dataset) and the `testing of hypotheses' (such as a 2-sample test on the equality of two medians).  

It is important that the scientist be aware of the range of applicability of a given inferential procedure. Several examples relevant to astronomical statistical practice can be mentioned.  Various statistics that resemble Pearson's $\chi^2$  do permit a weighted least squares regression, but the statistic often does not follow the $\chi^2$ distribution.  The Kolmogorov-Smirnov statistic can be a valuable measure of difference between a sample and a model, but tabulated probabilities are incorrect if the model is derived from that sample (Lilliefors 1969) and the statistic is ill-defined if the dataset is multivariate.  The likelihood ratio test can compare the ability of two models to explain a dataset, but it cannot be used if an estimated parameter value is consistent with zero (Protassov et al. 2002).

\subsection{Probability theory and statistical distributions}

Statistics is rooted in probability theory, a branch of mathematics seeking to model uncertainty (Ross 2010).  Nearly all astronomical studies encounter uncertainty: observed samples represent only small fractions of underlying populations, properties of astrophysical interest are measured indirectly or incompletely, errors are present due to the measurement process.   The theory starts with the concept of an `experiment', an action with various possible results where the actually occurring result cannot be predicted with certainty prior to the action.  Counting photons at a telescope from a luminous celestial object, or waiting for a gamma-ray burst in some distant galaxy are examples of experiments.  An `event' is a subset of the  (sometimes infinite) `sample space', the set of all outcomes of an experiment.  For example, the number of supermassive black holes within 10 Mpc is a discrete and finite sample space, while the spatial distribution of galaxies within 10 Mpc can be considered as an infinite sample space.   

Probability theory seeks to assign probabilities to elementary outcomes and manipulate the probabilities of elementary events to derive probabilities of complicated events.  Three `axioms of probability' are:  the probability $P(A)$ of an event $A$ lies between 0 and 1, the sum of probabilities over the sample space is 1, and the joint probability of two or more events is equal to the sum of individual event probabilities if the events are mutually exclusive.  Other properties of probabilities flow from these axioms: additivity and inclusion-exclusion properties, conditional and joint probabilities, and so forth.  

`Conditional probabilities' where some prior information is available are particularly important.  Consider an experiment with $m$ equally likely outcomes and let $A$ and $B$ be two events. Let $\#A = k$, $\#B = n$, and $\#(A \cap B) = i$ where $\cap$ means the intersection (`and').  Given information that $B$ has happened, the probability that $A$ has also happened is written $P(A \mid B) = i/n$, this is the conditional probability and is stated `The probability that $A$ has occurred given $B$ is $i/n$'.  Noting that $P(A\cap  B)=\frac{i}{m}$ and $P(B)=\frac{n}{m}$, then 
\begin{equation}
P(A\mid B)=\frac{P(A\cap  B)}{P(B)}.
\end{equation}
This leads to the `multiplicative rule' of probabilities, which for $n$ events can be written 
\begin{eqnarray}
P(A_1\cap A_2 \cap\ldots A_n) = & P(A_1)\, P(A_2\mid A_1) \, \dots \, P(A_{n-1}\mid A_1,\ldots A_{n-2})  \nonumber \\
 & \times P(A_n\mid A_1,\ldots A_{n-1}).
\end{eqnarray}
Let $B_1, B_2, \ldots, B_k$ be a partition of the sample space.   The probability of outcome $A$ in terms of events $B_k$, 
\begin{equation}
P(A) = P(A|B_1)P(B_1) + \cdots + P(A|B_k)P(B_k),
\end{equation}
is known as the `Law of Total Probability'.  The question can also be inverted to find the probability of an event $B_i$ given $A$,
\begin{equation}
P(B_i\mid A)\,=\,\frac{P(A\mid  B_i)P(B_i)}{P(A\mid  B_1)P(B_1) + \cdots + P(A\mid  B_k)P(B_k)}.
\end{equation}
This is known as `Bayes' Theorem' and, with a particular interpretation, it serves as the basis for Bayesian inference. 

Other important definitions and results of probability theory are important to statistical methodology.  Two events $A$ and $B$ are defined to be `independent' if $P(A\cap B)=P(A)P(B)$.  Random variables are functions of the sample or outcome space.  The `cumulative distribution function' (c.d.f.) $F$ of a random variable $X$ is defined as 
\begin{equation}
F(x) = P(X \le x).
\end{equation}
In the discrete case where $X$ takes on values $a_1, a_2, \ldots, a_n$, then $F$ is defined through the probability mass function (p.m.f.) $P(a) = P(x=a_i)$ and 
\begin{equation}
F(x) = \sum_{a_x \leq x} P(x=a_i).
\end{equation}
Continuous random variables care often described through a `probability density distribution' (p.d.f.) $f$ satisfying $f(y) \geq 0$ for all $y$ and 
\begin{equation}
F(x) = P(X \leq x) = \int_{-\infty}^x f(y)dy. 
\end{equation}
Rather than using the fundamental c.d.f.'s, astronomers have a tradition of using binned p.d.f.'s, grouping discrete data to form discontinuous functions displayed as histograms,  Although valuable for visualizing data, this is an ill-advised practice for statistical inference: arbitrary decisions must be made concerning binning method, and information is necessarily lost within bins.  Many excellent statistics can be computed directly from the c.d.f. using, for example, maximum likelihood estimation (MLE).

`Moments' of a random variable are obtained from integrals of the p.d.f. or weighted sums of the p.m.f.  The first moment, the `expectation' or mean, is defined by  
\begin{equation}
E[X] =  \mu = \int x f(x)dx 
\end{equation}
in the continuous case and $E[X] = \sum_a aP(X=a)$ in the discrete case.  The variance, the second moment, is defined by
\begin{equation}
Var[X] = E[(X-\mu)^2]. 
\end{equation}

A sequence of random variables $X_1, X_2, \ldots, X_n$ is called `independent and identically distributed (i.i.d.)' if
\begin{equation}
P(X_1 \leq a_1, X_2 \leq a_2, \ldots,  X_n \leq a_n) = P(X_1 \leq a_1) P(X_2 \leq a_2) \ldots P(X_n \leq a_n),
\end{equation}
for all $n$.  That is,  $X_1, X_2, \ldots, X_2$ all have the same c.d.f., and the events $P(X_i \leq _i)$ are independent for all $a_i$.   The `Law of Large Numbers' is a theorem stating that 
\begin{equation}
\frac{1}{n} \sum_{i=1}^n X_i \approx E[X] 
\end{equation} 
for large $n$ for a sequence of i.i.d. random variables.  

The continuous `normal distribution', or Gaussian distribution, is described by its p.d.f.  
\begin{equation}
\phi(x) = \frac{1}{\sqrt{2\pi}\sigma} \,
\exp\left\{-\frac{(x-\mu)^2}{2\sigma^2}\right\}. 
\label{normal-density.eqn}
\end{equation}
When $X$ has the Gaussian density in (\ref{normal-density.eqn}), then the first two moments are 
\begin{equation}
E(X)=\mu \quad Var(X) = \sigma^2.
\end{equation}   
The normal distribution, often designated $N(\mu,\sigma^2)$, is particularly important as the Central Limit Theorem states that the distribution of the sample mean of any i.i.d. random variable about its true mean approximately follows a normal.

The Poisson random variable $X$ has a discrete distribution with p.m.f.  
\begin{equation}
P(X=i) = \lambda^i e^{-\lambda}/i!
\end{equation}
for integer $i$.  For the `Poisson distribution', the mean and variance are equal,
\begin{equation}
E(X) = Var(X) = \lambda.
\end{equation}
If $X_1, X_2, \ldots, X_n$ are independent random variables with the Poisson distribution having  rate $\lambda$, then $(1/n) \sum X_i$ is the best, unbiased estimator of $\lambda$.  If $x_1, x_2, \ldots, x_n$ is a particular sample drawn from the Poisson distribution with rate $\lambda > 0$, then the sample mean $\bar{x}=\sum x_i / n$ is the best unbiased estimate for $\lambda$.  Here the $x_i$ are the realizations of the random variables $X_i$'s.   The difference of two Poisson variables, and the proportion of two Poisson variables, follow no simple known distribution and their estimation can be quite tricky.  This is important because astronomers often need to subtract background from Poisson signals, or compute ratios of two Poisson signals.  For faint or absent signals in a Poisson background, MLE and Bayesian approaches have been considered (Cowan 2006, Kashyap et al. 2010).   For Poisson proportions, the MLE is  biased and unstable for small $n$ and/or $p$ near 0 or 1, and other solutions are recommended (Brown et al. 2001).  If background subtraction is also present in a Poisson proportion, then a Bayesian approach is appropriate (Park et al. 2006).   

The power law distribution is particularly commonly used to model astronomical random variables.  Known in statistics as the Pareto distribution, the correctly normalized p.d.f. is 
\begin{equation}
f(x) = \frac{\alpha b^\alpha}{x^{\alpha+1}},
\end{equation}
for $x>b$.  The commonly used least-squares estimation of the shape parameter $\alpha$ and scale parameter $b$ from a binned dataset is known to be biased and inefficient even for large $n$.  This procedure is not recommended and the `minimum variance unbiased estimator' (MVUE) based on the MLE is preferred (Johnson et al. 1994)
\begin{eqnarray}
\alpha^*  &=& \left( 1 - \frac{2}{n} \right) \hat{\alpha}_{MLE}  ~~~ {\rm where}~~~ \hat{\alpha}_{MLE} = \frac{n}{\sum_{i-1}^n \ln (x_i/\hat b_{MLE})} \nonumber \\
b^* &=& \left(1 - \frac{1}{(n-1) \hat{\alpha}_{MLE}} \right) \hat{b}_{MLE} ~~~ {\rm where} ~~~ \hat{b}_{MLE} = x_{min}.
\label{pareto_est.eqn}
\end{eqnarray}

MLE and other recommended estimators for standard statistics of some dozens of distributions are summarized by Evans et al. (2000, also in Wikipedia) and are discussed comprehensively in volumes by Johnson et al. (1994).

\subsection{Point estimation}

Parameters of a distribution or a relationship between random variables are estimated using functions of the dataset. The mean and variance, for example, are parameters of the normal distribution.   Astronomers fit astrophysical models to data, such as a Keplerian elliptical orbit to the radial velocity variations of a star with an orbiting planet or the Navarro-Frenk-White distribution of Dark Matter in galaxies.   These laws also have parameters that determine the shape of the relationships.  

In statistics, the term `estimator' has a specific meaning.  The estimator $\hat\theta$ of $\theta$ is a function of the random sample that gives the value estimate when evaluated at the actual data, $\hat\theta= f_n(X_1, \dots, X_n)$, pronounced `theta-hat'. Such functions of random variables are also called `statistics'. Note that an estimator is a random variable because it depends on the observables that are random variables. For example, in the Gaussian case (equation \ref{normal-density}), the sample mean $\hat\mu=\bar{X} = \frac1n\sum\limits_{i=1}^n X_i$ and sample variance defined as  $\hat\sigma^2= S^2 = \frac {1}{n-1}\sum\limits_{i=1}^n (X_i - \bar{X})^2$ are estimators of $\mu$ and $\sigma^2$ respectively.  If the $X_i$'s are replaced by actual data, $x_1, x_2, \ldots, x_n$, then $\hat\mu = \frac1n\sum\limits_{i=1}^n x_i$ is a point estimate of $\mu$.  

Least squares (LS), method of moments, maximum likelihood estimation (MLE), and Bayesian methods are important and commonly used procedures in constructing estimates of the parameters.  Astronomers often refer to point estimation by `minimizing $\chi^2$'; this is an incorrect designation and cannot be found in any statistics text.  The astronomers' procedure is a weighted LS procedure, often using measurement errors for the weighting, that, under certain conditions, will give a statistic that follow a $\chi^2$ distribution.   

The choice of estimation method is not obvious, but can be guided by the scientific goal.  A procedure that gives the closest estimate to the true parameter value (smallest bias) will often differ from a procedure that minimizes the average distance between the data and the model (smallest variance), the most probable estimate (maximum likelihood), or estimator most consistent with prior knowledge (Bayesian).   Astronomers are advised to refine their scientific questions to choose an estimation method, or compare results from several methods to see how the results differ.  

Statisticians have a number of criteria for assessing the quality of an estimator including unbiasedness, consistency, efficiency.  An estimator $\hat\theta$ of a parameter $\theta$ is called `unbiased' if the expected value of $\hat\theta$, $E(\hat\theta)=\theta$. That is, $\hat\theta$ is unbiased if its overall average value for all potential datasets is equal to $\theta$.  For example, in the variance estimator $S^2$ of $\sigma^2$ for the Gaussian distribution, $n-1$ is placed in the denominator instead of $n$ to obtain an unbiased estimator.  If $\hat\theta$ is unbiased estimator of $\theta$, then the variance of the estimator $\hat\theta$ is given by  $E((\hat\theta-\theta)^2)$.  Sometimes, the scientist will choose a minimum variance estimator to be the `best' estimator, but often a bias is accepted so that the sum of the variance and the square of the bias, or `mean square error' (MSE), is minimized,
\begin{equation}
MSE = E[(\hat\theta-\theta)^2] = Var(\hat\theta) + (\theta-E[\hat\theta])^2.
\end{equation}
An unbiased estimator that minimizes the MSE is called the `minimum variance unbiased estimator' (MVUE).   If there are two or more unbiased estimators,  the one with smaller variance is usually preferred. Under some regularity conditions, the `Cram\'er-Rao inequality' gives a lower bound on the lowest possible variance for an unbiased estimator. 

\subsection{Least squares}

The LS method, developed for astronomical applications two hundred years ago (\S 1), is effective in the regression context for general linear models.  Here `linear' means linear in the parameters, not in the variables.  LS estimation is thus appropriate for a wide range of complicated astrophysical models.  Suppose  $X_i$ are independent but not identically distributed, say, $E(X_i) = \sum_{j=1}^k a_{ij}\beta_j$, (the mean of $X_i$ is a known linear combination of parameters $\beta_1, \dots, \beta_k$). Then the estimators of the parameters $\beta_j$ can be obtained by minimizing the sum of squares of $(X_i - \sum_{j=1}^k a_{ij}\beta_j$). In some cases, there is a closed form expression for the least squares estimators of $\beta_1, \dots, \beta_k$. If the error variances $\sigma_i^2$ of $X_i$ are also different (heteroscadastic), the one can minimize the weighted sum of squares 
\begin{equation}
\sum_{i=1}^n \frac1{\sigma_i^2}\left(X_i - \sum_{j=1}^k a_{ij}\beta_j\right)^2
\end{equation}
over $\beta_1, \dots, \beta_k$.  If $Y_1, Y_2, \ldots, Y_n$ are independent random variables with  a normal distribution $N(\mu,1)$, then the sum of squared normals $\sum_{i-1}^n$ is distributed as the $\chi^2$ distribution.  

The method of moments is another classical approach to point estimation where the parameters of the model are expressed as simple functions of the first few moments and then replace the population moments in the functions with the corresponding sample moments.

\subsection{Maximum likelihood method}

Building on his criticism of  both least-squares method and the method of moments in  his  first mathematical paper as an undergraduate,  R. A. Fisher  (1922) introduced the method of maximum likelihood.  The method is based on  the `likelihood' where the p.d.f. (or probability mass function for a discrete random variable) is viewed as a function of the data given the model and specified values of the parameters. In statistical parlance, if the data are an i.i.d. random sample $X_1,\ldots,X_n$,  with a common p.d.f. or p.m.f. $f(., \theta)$, then the likelihood $L$ and loglikelihood $\ell$ are given by 
\begin{equation}
\ell(\theta) = \ln L(\theta) = \sum_{i=1}^n \, \ln f(X_i, \theta).
\end{equation}
In Fisher's formulation, the model parameters $\theta$ are treated as fixed and the data are variable. 

The `maximum likelihood estimator' (MLE) $\hat\theta$ of $\theta$ is the value of the parameter that maximizes $\ell(\theta) $. While this can be calculated analytically for simple models,  MLE estimators can often be numerically calculated for more complex functions of the data.  For many useful interesting functions $g$ of the parameters $\theta$, $g(\hat\theta)$ is the MLE of $g(\theta)$ whenever $\hat\theta$ is the MLE of $\theta$.  Computing the maximum likelihood is usually straightforward.  The `EM Algorithm' is an easily implemented and widely used procedure for maximizing likelihoods  (Dempster et al. 1977, McLachlan \& Krishnan 2008).  This procedure, like many other optimization calculations, may converge to a local rather than the global maximum.Astronomers use the EM Algorithm for an MLE in image processing where it is called the Lucy-Richardson algorithm (Lucy 1974).    In some cases, the MLE may not exist and, in other cases, more than one MLEs exist. 

Knowledge of the limiting distribution of the estimator is often needed to obtain confidence intervals for parameters. In many  commonly occurring situations with large $n$, the MLE $\hat\theta$ has an approximate normal distribution with mean $\theta$ and variance $1/I(\theta)$ where 
\begin{equation}
I(\theta) = n E\left(\frac{\partial}{\partial\theta} \log f(X_1,\theta)\right)^2.
\label{FisherInfo}
\end{equation}
This is the Fisher information matrix.  Thus, 95\% (or similar) confidence intervals can be derived for MLEs.  

With broad applicability, efficient computation, clear confidence intervals, and strong mathematical foundation, maximum likelihood estimation rose to be the dominant method for parametric estimation in many fields.  Least squares methodology still predominates in astronomy, but MLE has a growing role.  An important example in astronomy of MLE with Fisher information confidence intervals was the evaluation of cosmological parameters of the concordance $\Lambda$ Cold Dark Matter based on fluctuations of the cosmic microwave background radiation measured with the Wilkinson Microwave Anisotropy Probe (Spergel et al 2003).

\subsection{Hypotheses tests}

Along with estimation, hypotheses tests are a major class of tools in statistical inference.  Many astronomical problems such as source detection and sample comparison can be formulated as Yes/No questions to be addressed with statistical hypotheses testing.

Consider the case of source detection where the observed data $Y$ consists of signal $\mu$ and noise $\epsilon$,  $Y=\mu + \epsilon$. The problem is to quantitatively test the `null hypothesis' $H_0: \mu = 0$ representing no signal  against the `alternative hypothesis' $H_a: \mu > 0$. Statistical hypothesis testing resembles a court room trial where a defendant is considered innocent until proven guilty.  A suitable function of the data called the `test statistic' is chosen, and a set of test statistic values or  `critical region' is devised.  The decision rule is to reject the null hypothesis if the function of the data falls in the critical region. There are two possible errors:  a false positive that rejects the null hypothesis, called a `Type I error'; and a false negative that fails to reject the null hypothesis when the alternative hypothesis is true, or   `Type II error'.  Ideally, one likes to minimize both error types, but it is impossible to achieve.  Critical regions constructed to keep Type I error under control, say at 5\% level are called `levels of significance'.  One minus the probability of Type II error is called the `power of the test'.  High power tests are preferred.  

A result of a hypothesis test is called `statistically significant' if it is unlikely to have occurred by chance, that is, the test rejects the null hypothesis at the prescribed significance level $\alpha$ where $\alpha = 0.05$, 0.01 or similar value.  Along with the results of a statistical test, often the so-called $p$-value is reported. The `$p$-value' of the test is the smallest significance level at which the statistic is significant.  It should be important to note that the null hypothesis and the alternative hypothesis are not treated symmetrically: the null hypothesis can be rejected at a given level of significance, but the null hypothesis can not formally be accepted.

\subsection{Bayesian estimation}

Conceptually,  Bayesian inference uses aspects of the scientific method that involves evaluating whether acquired evidence is consistent or inconsistent with a given hypothesis. As evidence accumulates, the degree of belief in a hypothesis ought to change. With enough evidence, it should become very high or very low. Thus,  Bayesian inference can be used to discriminate between conflicting hypotheses: hypotheses with very high support should be accepted as true and those with very low support should be rejected as false. However, this inference method is influenced by the prior distribution, initial beliefs that one holds before any evidence is ever collected.  In so far as the priors are not correct, the estimation process can lead to false conclusions.  

Bayesian inference relies on the concept of conditional probability to revise one's
knowledge. Prior to the collection of sample data one had some (perhaps vague)
information on $\theta$.  Then combining the model density of the observed data with
the prior density one gets the posterior density, the conditional density of $\theta$ given
the data. Until further data is available, this posterior distribution of $\theta$ is the only
relevant information as far as is concerned.

As outlined above, the main ingredients for Bayesian inference are the likelihood function, $L(\theta\mid x)$ with a vector $\theta$ of parameters and a prior probability density, $\pi(\theta)$. Combining the two via Bayes theorem yields the posterior probability density
\begin{equation}
\pi(\theta \mid x) = \frac{\pi(\theta) ~ L(\theta \mid x)} {\int \pi(u) ~ L(u \mid x) du}
\end{equation}
when the densities exist.  In the discrete case where $\pi(\theta)$ is the probability mass function, the formula becomes
\begin{equation}
\pi(\theta \mid x) = \frac{\pi(\theta) ~ L(\theta \mid x)} {\sum_{j=2}^k  \pi(\theta_j) ~ L(\theta_j \mid x)}
\end{equation}

If there is no special information on the parameter $\theta$ except that it lies in an interval, then one often assumes $\theta$ is uniformly distributed on the interval. This is a choice of a `non-informative prior' or reference prior. Often, Bayesian inference from such a flat prior coincides with classical frequentist inference.

The estimator $\hat\theta$ of $\theta$ defined as the mode of $\pi(\theta | x)$, the value of $\theta$ that maximizes the posterior $\pi((\theta | x)$,  is the most probable value of the unknown parameter $\theta$ conditional on the sample data.  This is called the `maximum a posteriori' (MAP) estimate or the `highest posterior density' (HPD) estimate.

The mean of the posterior distribution gives another Bayes estimate by applying least squares on the posterior density. Here the $\hat\theta_B$ that minimizes the posterior dispersion
\begin{equation}
E[(\theta - \hat\theta_B)^2 \mid x] = {\rm min}~ E[(\theta -a)^2 \mid x]
\end{equation}
is given by $\hat\theta_B=E[\theta \mid x]$.  If $\hat\theta_B$ is chosen as the estimate of $\theta$, then a measure of variability of this estimate is the posterior variance, $E[(\theta - E[\theta \mid x])^2 |x|$.  This gives the posterior standard deviation as a natural measure of estimation error; that is,  the estimate is $\hat\theta_B \pm \sqrt{E[(\theta - E[\theta \mid x])^2 |x|}$.   In fact, for any interval around $\hat\theta_B$,  the posterior probability containing the true parameter can be computed.  In other words, a statement such as
\begin{equation}
P(\hat\theta_B - k_1 \leq \theta \leq \hat\theta_B + k_2 | x) = 0.95
\end{equation}
gives a meaningful 95\% `credible region'.  These inferences are all conditional on the given dataset $x$.

Bayesian inference can be technically challenging because it requires investigating the full parameter space.  For models with many parameters and complex likelihood functions, this can involve millions of calculations of the likelihood. Sophisticated numerical methods to efficiently cover the parameter space are needed, most prominently using Markov chains with the Gibbs sampler and Metropolis-Hastings algorithm.  These are collectively called `Markov Chain Monte Carlo' calculations. Once best-fit $\hat\theta$ values have been identified, further Monte Carlo calculations can be performed to examine the posterior distribution around the best model.  These give credible regions in parameter space.  Parameters of low scientific interest can be integrated to remove them from credible region calculations, this is called `marginalization' of nuisance parameters.  

The Bayesian framework can also effectively choose between two models with different vectors of parameters. This is the `model selection' problem discussed in \S\ref{modelselection.sec}.  Suppose the data $X$ has probability density function $f(x \mid \theta)$ and the scientist wants to compare two models, $M_0 : \theta \in \Theta_0$ $vs.$ $M_1 : \theta \in \Theta_1$.  A prior density that assigns positive prior probability to $\Theta_0$ and $\Theta_1$ are chosen, and the posterior `odds ratio' $P\{\Theta_0|x\}/P\{\Theta_1|x\}$  is calculated.  A chosen threshold like 1/9 or 1/19 will decide what constitutes evidence against a null hypothesis.  The `Bayes factor' of $M_0$ relative to $M_1$ can also be reported,
\begin{equation}
BF_{01} = \frac{P(\Theta_0|x)}{P(\Theta_1|x)} / \frac{P(\Theta_0)}{P(\Theta_1)}
= \frac{\int_{\Theta_0} f(x|\theta)g_0(\theta) d\theta}{\int_{\Theta_1} f(x|\theta)g_1(\theta) d\theta}.
\end{equation}
The smaller the the value of $BF_{01}$, the stronger the evidence against $M_0$.  Unlike classical hypothesis testing, the Bayesian analysis treats the hypotheses symmetrically.  The method can be extended to compare more than two models.

\subsection{Resampling methods}

Astronomers often devise a statistic that measures a property of interest in the data, but find it is difficult or impossible to determine the distribution of that statistic.  The classical statistical methods concentrate on statistical properties of estimators that have a simple closed form, but these methods often involve unrealistically simplistic model assumptions. A class of computationally intensive procedures known as `resampling methods' address this limitation, providing inference on a wide range of statistics under very general conditions.  Resampling methods involve constructing hypothetical datasets derived from the observations, each of which can be analyzed in the same fashion to see how the chosen statistic depend on plausible random variations in the observations.  Resampling the original data preserves whatever distributions are truly present, including selection effects such as truncation and censoring.

The `half-sample method' is an old resampling method dating to the 1940s.  Here  one repeatedly chooses at random half of the data point, and estimates the statistic for each resample.  The inference on the parameter can be based on the histogram of the resampled statistics.  An important variant is the Quenouille--Tukey `jackknife method' where one constructs exactly $n$ hypothetical datasets each with $n - 1$ points, each one omitting a different point.  It is useful in reducing the bias of an estimator as well as estimating the variance of an estimator.  The jackknife method is effective for many statistics, including LS estimators and MLEs, but is not consistent for discrete statistics such as the sample median.  

The most important of resampling methods is the `bootstrap' introduced by Bradley Efron in 1979 (Efron \& Tabishrani 1993).   Here one generates a large number of datasets, each randomly drawn from the original data such that each drawing is made from the entire dataset, so a simulated dataset is likely to miss some points and have duplicates or triplicates of others. This `resampling with replacement' can be viewed as a Monte Carlo simulation from an existing data without any assumption on the underlying population.

The importance of the bootstrap emerged during the 1980s when mathematical study demonstrated that it gives nearly optimal estimate of the distribution of many statistics under a wide range of circumstances (Babu \& Singh 1983, Babu 1984).   For example, theorems using Edgeworth expansions establish that the bootstrap provides a good approximation for a Studentized smooth functional model (Babu \& Singh 1984).   A broad class of common statistics can be expressed as smooth functions of multivariate means including LS estimators (means and variances, $t$-statistics, correlation coefficients, regression coefficients) and some MLEs.  The bootstrap is consequently widely used for a vast range of estimation problems. 

While bootstrap estimators have very broad application, they can fail for statistics with heavy tails, some non-smooth and nonlinear situations, and some situations where the data are not independent.  A lack of independence can occur, for example, in proximate pixels of an astronomical image due to the telescope point spread function, or in proximate observations of a time series of a variable celestial object.  The bootstrap may also be inapplicable when the data have `heteroscedastic' measurement errors; that is, the variances that differ from point to point.   Bootstrap confidence intervals also require that the statistic be `pivotal' such that the limiting distribution is free from the unknown parameters of the model.  Fortunately, methods are available to construct approximately pivotal quantities in many cases, and in the dependent case such as autocorrelated images or time series, a modification called the `block bootstrap' can be applied.  Loh (2008) describes an application to the galaxy two-point correlation function.  

The most popular and simple bootstrap is the `nonparametric bootstrap' where the resampling with replacement is based on the `empirical distribution function' (e.d.f.) of the original data.   The `parametric bootstrap' uses a functional approximation, often a LS or MLE fit, rather than the actual dataset to obtain random points.  This is a well-known simulation procedure (Press et al. 1986).   Bootstrapping a regression problem requires a choice:  one can bootstrap the residuals from the best fit function (classical bootstrap),  or one can bootstrap multivariate data points (paired bootstrap).  The paired bootstrap is robust against heteroscadasticity in the errors.

\subsection{Model selection and goodness of fit} 
\label{modelselection.sec}
 
The aim of model fitting is to provide most parsimonious ``best" fit of a parametric model to data.  It might be a simple heuristic model to phenomenological relationships between observed properties in a sample of astronomical objects, or a more complex model based on astrophysical theory. A good statistical model should be parsimonious yet conforming to  the data, following the principle of Occam's Razor. A satisfactory model avoids underfitting which induces bias, and avoids overfitting which induces high variability.   A model selection criterion should balance the competing objectives of conformity to the data and parsimony.  

The statistical procedures for parameter estimation outlined above, such as LS and MLE, can link data with astrophysical models, but they do not by themselves evaluate whether the chosen model is appropriate for the dataset.  The relevant methods fall under the rubrics of statistical `model selection', and `goodness of fit'.  The common procedure in astronomy based on the reduced chi-squared $\chi^2_\nu \simeq 1$ is a primitive technique not used by statisticians or researchers in other fields.  

Hypothesis testing discussed above can be used to compare two models which share a common structure and some parameters, these are `nested models'.  However, it does not treat models symmetrically.  A more general framework for model selection will be based on likelihoods.  Let $D$ denote the observed data and $M_1,\dots,M_k$ denote models for $D$ under consideration.  For each model $M_j$, let $f(D|\theta_j,M_j)$ denotes the likelihood, the p.d.f. (or p.m.f. in the discrete case) evaluated at the data $D$, and let $\ell(\theta_j)=\log f(D|\theta_j,M_j)$ denote the  loglikelihood  where $\theta_i$ is a $p_j$ dimensional parameter vector.  

Three classical hypothesis tests based on MLEs for comparing two models were developed during the 1940s.  To test the null hypothesis $H_0: \theta = \theta_0$, the Wald Test uses $W_n= (\hat\theta_n - \theta_0)^2/Var(\hat\theta_n)$, the standardized distance between $\theta_0$ and the maximum likelihood estimator $\hat\theta_n$ based on a dataset of size $n$. The distribution of $W_n$ is approximately chi-square with one degree of freedom. In general, the variance of $\hat\theta_n$ is not known, however, a close approximation is $1/I(\hat\theta_n)$, where $I(\theta)$ is the Fisher's information. Thus $I(\hat\theta_n)(\hat\theta_n - \theta_0)^2$ has a chi-square distribution in the limit, and the Wald test rejects the null hypothesis $H_0$, when this quantity is large.  The `likelihood ratio test' uses the logarithm of ratio of likelihoods, $\ell(\hat\theta_n)-\ell(\theta_0)$, and Rao's score test uses the statistic $S(\theta_0) =(\ell^\prime(\theta_0))^2/(nI(\theta_0))$, where $\ell^\prime$ denotes the derivative of $\ell$.   The likelihood ratio is most commonly used in astronomy.  Protassov et al. (2002) warn about its common misuse.  

If the model $M_1$ happens to be nested in the model $M_2$, the largest likelihood achievable by $M_2$ will always be larger than that achievable by $M_1$.   This suggests that the addition of a penalty on models with more parameters would achieve a balance between overfitting and underfitting. Several penalized likelhood approaches to model selection have been actively used since the 1980s.  The `Akaike's Information Criterion' (AIC), based on the concept of entropy, for model $M_j$ is defined to be 
\begin{equation}
AIC = 2\ell(\hat\theta_j) - 2p_j.
\label{aic.eqn}
\end{equation}  
Unlike hypothesis tests, the AIC does not require the assumption that one of the candidate models is correct, it treats models symmetrically, and can compare both nested and non-nested models.  Disadvantages of the AIC include the requirement of large samples and the lack of consistency in giving the true number of model parameters even for very large $n$.    The `Bayesian Information Criterion' (BIC) is a popular alternative model selection criterion defined to be 
\begin{equation}
BIC = 2\ell (\hat{\theta}_j)  - p_j\log n.
\end{equation}  
Founded in Bayesian theory, it is consistent for large $n$.  The AIC penalizes free parameters less strongly than does the BIC. A more difficult problem is comparing best-fit models derived for non-nested model families.   One possibility is using the `Kullback-Leibler information', a measure of proximity between data and model arising from information theory.   

Goodness of fit can be estimated using nonparametric tests similar to the Kolmogorov-Smirnov statistic discussed in \S\ref{nonparametric.sec}.  However, the goodness of fit probabilities derived from these statistics are usually not correct when applied in model fitting situations when the parameters are estimated from the dataset under study. An appropriate approach is bootstrap resampling that gives valid estimates of goodness of fit probabilities under a very wide range of situations.  Both the nonparametric and parametric bootstrap can be applied for goodness of fit tests.  The method cannot be used for multivariate data due to identifiability problems.

\section{Applied fields of statistics}

\subsection{Nonparametric statistics}
\label{nonparametric.sec}

Nonparametric statistical inference gives insights into data which do not depend on  assumptions regarding the distribution of the underlying population.  Most standard statistics implicitly assume, through the Central Limit Theorem, that all distributions are normal, measurement uncertainties are constant and increase as $\sqrt{N}$ as the sample size increases, and chosen parametric models are true.  But our knowledge of astronomical populations and processes $-$ Kuiper Belt Objects,  Galactic halo stellar motions, starburst galaxies properties, and accretion onto supermassive black holes, and so forth $-$  is very limited.  The astronomer really does not know that the observed properties using convenient units are in fact normally distributed or that relationships between properties are in fact (say) power law.  Nonparametric approaches to statistical inference should thus be particularly attractive to astronomers, and can precede more restrictive parametric analysis.

Nonparametric statistics are called `distribution-free' because they are valid for any underlying distribution and any  transformation of the variables.  Some methods are particularly `robust' against highly skewed distributions or outliers due to erroneous measurements or extraneous objects.  Some are based on the rankings of each object within the dataset.  However, many nonparametric methods are restricted to univariate datasets; for example, there is no unique ranking for a bivariate dataset.  

Nonparametric analysis often begins with `exploratory data analysis' as promoted by statistician John Tukey.  The `boxplot' is a compact and informative visualization of a univariate dataset (Figure~\ref{PSSS_Feigelson_univar.fig}).  It displays the five-number summary (minimum, 25\% quartile, median, 75\% quartile, maximum) with whiskers, notches and outliers.  There is a broad consensus that the `median', or central value, of a dataset is the most reliable measure of location.  `Trimmed means' are also used. The spread around the median can be evaluated with the `median absolute deviation' (MAD) that, when normalized to the normal standard deviation, is given by 
\begin{equation}
MAD(x) = 1.483 \times {\rm Median}~|x_i - {\rm Median}(x)|
\end{equation}

The cumulative distribution of a univariate distribution is best estimated by the `empirical distribution function' (e.d.f.), 
\begin{equation}
\hat{F}_n(x) = \frac{1}{n} \sum_{i=1}^n I[x_i \leq x]
\end{equation}
where $I$ is the indicator function.  It ranges from 0.0 to 1.0 with step heights of $1/n$ at each observed value.  For a significance level $\alpha = 0.05$ or similar value, the approximate confidence interval for $F(x)$ ,the true distribution function at $x$, is given by
\begin{equation}
\widehat{F}_n(x) \pm z_{1-\alpha/2} \sqrt{\widehat{F}_n(x)[1-\widehat{F}_n(x)]/n}
\end{equation}
where $z_\alpha$ are the quantiles of the Gaussian distribution. 

Important nonparametric statistics are available to test the equality of two e.d.f.'s, or for the compatibility of an e.d.f. with a model.  The three main statistics are:
\begin{eqnarray}
{\rm Kolmogorov-Smirnov ~(KS)} && M_{KS}=\max_{x} | \widehat{F}_n(x) -F_0(x) | \nonumber \\
{\rm Cramer-von~Mises ~(CvM)} && W_{CvM}^2 = n \sum_{i-1}^n (\widehat{F}_n(x_i) - F_0(x_i))^2 \nonumber \\
{\rm Anderson-Darling ~(AD)} && A_{AD}^2 = n \sum_{i=1}^n \frac{(\widehat{F}_n(x_i) - F_0(x_i))^2}{F_0(x_i)(1-F_0(x_i)}.
\end{eqnarray}
The KS statistic is most sensitive to large-scale differences in location (i.e., the median value) and shape between the two distributions.  The CvM statistic is effective for both large-scale and small-scale differences in distribution shape.  But both of these measures are relatively insensitive to differences near the ends of the distribution. This deficiency is addressed by the AD statistic, a weighted version of the C-vM statistic to emphasize differences near the ends.  The AD test is demonstrably the most sensitive of the e.d.f. tests; this was confirmed in a recent astronomical study by Hou et al. (2009).   The distributions of these statistics are known and are  distribution-free for all continuous $F$.  
 
But all these statistics are no longer distribution-free under two important and common situations: when the data are multivariate, or when the model parameters are estimated using the dataset under study.  Although astronomers sometimes use two-dimensional KS-type tests, these procedures are not mathematically validated to be distribution-free.  Similarly, when comparing a dataset to a model, the e.d.f. probabilities are distribution-free only if the model is fully specified independently of the dataset under study.  Standard tables of e.d.f probabilities thus do not give a mathematically correct goodness of fit test.  Fortunately, a simple solution is available: the distribution of the e.d.f. statistic can be established for each dataset using bootstrap resampling.  Thus, a recommended nonparametric goodness of fit procedure combines the sensitive Anderson-Darling statistic with bootstrap resampling to establish its distribution and associated probabilities.

\subsection{Data smoothing}

Density estimation procedures smooth sets of individual measurements into continuous curves or surfaces.  `Nonparametric density estimation' makes no assumption regarding the underlying distribution.  A common procedure in astronomy is to collect univariate data into histograms giving frequencies of occurrences grouped into bins.  While useful for exploratory examination, statisticians rarely use histograms for statistical inference for several reasons.  The choice of bin origin and bin width is arbitrary, information is unnecessarily lost within the bin, the choice of bin center is not obvious, multivariate histograms are difficult to interpret, and the discontinuities between bins does not reflect the continuous behaviors of most physical quantities. 

`Kernel density estimation', a convolution with a simple unimodal kernel function,  avoids most of these disadvantages and is a preferred method for data smoothing.  For an i.i.d. dataset, either univariate or multivariate, the kernel estimator is 
\begin{equation}
\hat{f}_{kern}(x,h) =  \frac{1}{nh(x)} \sum_{i=1}^n K \left( \frac{x - x_i}{h(x)} \right)
\label{kde.eqn}
\end{equation} 
where $h(x)$ is the `bandwidth' and the kernel function $K$ is normalized to unity.   The kernel shape is usually chosen to be a Gaussian or Epanechikov (inverted parabola) function.  Confidence intervals for $f(x)$ for each $x$ can be readily calculated around the smoothed distribution, either by assuming asymptotic normality if the sample is large or by bootstrap resampling.

The choice of bandwidth is the greatest challenge.  Too large a bandwidth causes oversmoothing and increases bias, while too small a bandwidth causes undersmoothing and increases variance.  The usual criterion is to choose the bandwidth to minimize the `mean integrated square error' (MISE),
\begin{equation}
MISE(\hat{f}_{kern})=  E[(\hat{f}_{kern} (x) - f(x))^2].
\end{equation}

A heuristic bandwidth for unimodal distributions, known as Silverman's rule-of-thumb, is $h = 0.9 \sigma n^{-1/5}$ where $\sigma$ is the standard deviation of the variable and $n$ is the number of data points.  A more formal approach is `cross-validation' which maximizes the log-likelihood of estimators obtained from jackknife simulations.  A variety of adaptive smoothers are used where $h(x)$ depends on the local density of data points, although there is no consensus on a single optimal procedure.  One simple option is to scale a global bandwidth by the local estimator value according to $h(x_i) = h / \sqrt{\hat{f}(x_i)}$.  An important bivariate local smoother is the `Nadaraya-Watson estimator'.  Other procedures are based on the distance to the `$k$-th nearest neighbor' ($k$-nn) of each point; one of these is applied to an astronomical dataset in Figure~\ref{PSSS_Feigelson_adapt.fig}.    

A powerful suite of smoothing methods have recently emerged known as `semi-parametric regression' or `nonparametric regression'.  The most well-known variant is William Cleveland's LOESS method that fits polynomial splines locally along the curve or surface.  Extensions include local bandwidth estimation from cross-validation, projection pursuit and kriging.  Importantly, numerically intensive calculations in these methods give confidence bands around the estimators.  These methods have been introduced to astronomy by Miller et al. (2002)  and Wang et al. (2005).

\subsection{Multivariate clustering and classification}

Many astronomical studies seek insights from a table consisting of measured or inferred properties (columns) for a sample of celestial objects (rows).  These are multivariate datasets.  If the population is homogeneous, their structure is investigated with methods from multivariate analysis such as principal components analysis and multiple regression.   But often the discovery techniques capture a mixture of astronomical classes.  Multi-epoch optical surveys such as the planned Large Synoptic Survey Telescope (LSST) will find pulsating stars, stellar eclipses from binary or planetary companions, moving asteroids, active galactic nuclei, and explosions such as novae, supernovae, and gamma-ray bursts.  Subclassifications are common.  Spiral galaxy morphologies were divided into Sa, Sb and Sc categories by Hubble, and later were given designations like SBab(rs).  Supernovae were divided into Types Ia, Ib and II, and more subclasses are considered.  

However, astronomers generally developed these classifications in a heuristic manner with informed but subjective decisions, often based on visual examinations of two-dimensional projections of the multivariate datasets.  A popular method for clustering in low-dimensions is the `friends-of-friends algorithm', known in statistics as single linkage hierarchical clustering.  But many astronomers are not aware that this clustering procedure has serious deficiencies and many alternatives are available. As an astronomical field matures, classes are often defined from small samples of well-studied prototypes that can serve as `training sets' for supervised classification.  The methodologies of unsupervised clustering and supervised classification are presented in detail by Everitt et al. (2001), Hastie et al. (2001) and Duda et al. (2002).  Many of the classification procedures have been developed in the computer science, rather than statistics, community under the rubrics of `machine learning' and `data mining'.  

As with density estimation, astronomers often seek nonparametric clustering and classification as there is no reason to believe that stars, galaxies and other classes have multivariate normal (MVN)  distributions in the observed variables and units.  However, most nonparametric methods are not rooted in probability theory, the resulting clusters and classes can be very sensitive to the mathematical procedure chosen for the calculation, and it is difficult to evaluate statistical significance for purported structures.   Trials with different methods, bootstrap resampling for validation, and caution in interpretation are advised. 

Most clustering and classification methods rely on a metric that defines distances in the $p$-space, where $p$ is the number of variables or `dimensionality' of the dataset.  A Euclidean distance (or its generalization, a Minkowski $m$-norm distance) is most often adopted, but the distances then depend on the chosen units that are often incompatible (e.g., units in a stellar astrometric catalog may be in degrees, parsecs, milliarcsecond per year, and kilometers per second).  A common solution in statistics is to standardize the variables, 
\begin{equation}
x_{std} = \frac{x-\bar{x}}{\sqrt{Var(x)}}
\end{equation} 
where the denominator is the standard deviation of the dataset.  Astronomers typically choose a logarithmic transformation to reduce range and remove units.  A second choice needed for most clustering and classification methods is the definition of the center of a group.  Centroids (multivariate means) are often chosen, although medoids (multivariate medians) are more robust to outliers and classification errors.   A third aspect of a supervised classification procedure is to quantify classificatory success with some combination of Type 1 errors (correct class is rejected) and Type II errors (incorrect class is assigned). 

Unsupervised agglomerative hierarchical clustering is an attractive technique for investigating the structure of a multivariate dataset.  The procedure starts with $n$ clusters each with one member. The clusters with the smallest value in the pairwise `distance matrix' are merged, their rows and columns are removed and replaced with a new row and column based on the center of the cluster. This merging procedure is repeated $n$ times until the entire dataset of $n$ points is contained in a single cluster. The result is plotted as a Ôclassification treeÕ or ÔdendrogramÕ.  The structure depends strongly on the definition of the distance between a cluster and an external data point.  In single linkage clustering, commonly used by astronomers, the nearest point in a cluster is used.  However, in noisy or sparse data, this leads to spurious `chaining' of groups into elongated structures.  For this reason, single linkage is discouraged by statisticians, although it may be appropriate in the search for filamentary patterns.   Average linkage and Ward's minimum variance method give a good compromise between elongated and hyperspherical clusters.  

`$k$-means partitioning' is another widely used method that minimizes the sum of within-cluster squared distances.  It is related both to Voronoi tesselations and to classical MANOVA methods that rely on the assumption of MVN clusters.    $k$-means calculations are computationally efficient as the distance matrix is not  calculated and cluster centroids are easily updated as objects enter or depart from a cluster.  A limitation is that the scientist must choose in advance the number $k$ of clusters present in the dataset.  Variants of $k$-means, such as robust $k$-medoids and the Linde-Buzo-Gray algorithm, are widely used in computer science for pattern recognition in speech and for image processing or compression.    

MLE clustering based on the assumption of MVN structures are also used with model selection (i.e., choice of number of clusters in the best model) using the Bayesian Information Criterion.  This is an implementation of `normal mixture models' and uses the `EM Algorithm' for maximizing the likelihood.  Other methods, such as DBSCAN and BIRCH, have been recently developed by computer scientists to treat more difficult situations like clusters embedded in noise, adaptive clustering and fragmentation, and efficient clustering of megadatasets. 

Techniques for supervised classification began in the 1930s with Fisher's `linear discriminant analysis' (LDA). For two classes in a training set, this can be viewed geometrically as the projection of the cloud of $p$-dimensional points onto a 1-dimensional line that maximally separates the classes.  The resulting rule is applied to members of the unclassified test set.  LDA is similar to principal components analysis but with a different purpose: principal components finds linear combinations of the variables that sequentially explain variance for the sample treated as a whole, while LDA finds linear combinations that efficiently separate classes within the sample.  LDA can be formulated with likelihoods for MLE and Bayesian analysis, and has many generalizations.  These include the important classes of machine learning techniques including `Support Vector Machines'.  

`Nearest neighbor classifiers'  ($k$-nn) are a useful class of techniques whereby cluster membership of a new object is determined by a vote among the memberships of the $k$ nearest neighboring points in the training set.  As in kernel density estimation, the choice of $k$ balances bias and variance, and can be made using cross-validation.  Bootstrap resampling can assist in evaluating the stability of cluster number and memberships.  In `discriminant adaptive nearest neighbor' classification, the metric is adjusted to the local density of points, allowing discovery of subclusters in dense regions without fragmenting low density regions. 

Astronomers often define classes by rules involving single variables, such as `Class III pre-main sequence stars have mid-infrared spectral indices $ [5.8]-[8]  < 0.3$' or `Short gamma-ray bursts have durations $< 2$ seconds'.  These partition the datasets along hyperplanes parallel to an axis, although sometimes oblique hyperplanes are used.  These criteria are usually established heuristically by examination of bivariate scatterplots, and no guidance is provided to estimate the number of classes present in the dataset.  In statistical parlance, these rule-based classifications are `classification trees'.   Mature methodologies called  `classification and regression trees' (CART) have been developed by Leo Breiman and colleagues to grow, prune, and evaluate the tree.  Sophisticated variants like bootstrap aggregation (`bagging') to quantify the importance and reliability of each split and `boosting' to combine weak classification criteria, have proved very effective in improving CART and other classification procedures.  Important methods implementing these ideas include AdaBoost and Random Forests.  CART-like methods are widely used in other fields and could considerably help astronomers with rule-based classification.

Many other multivariate classifiers are available for both simple and complex problems: naive Bayes, neural networks, and so forth.  Unfortunately, as most methods are not rooted in mathematical statistics, establishing probabilities for a given cluster or pattern in a training set, or probabilities for assigning new objects to a class, are difficult or impossible to establish.  Indeed, a formal `No Free Lunch Theorem' has been proved showing that no single machine learning algorithm can be demonstrated to be better than another in the absence of prior knowledge about the problem.  Thus, the astronomer's subjective judgments will always be present in multivariate classification; however, these can be informed by quantitative methodologies which are not yet in common use.

\subsection{Nondetections and truncation}

Astronomical observations are often subject to selection biases due to limitations of the telescopes.  A common example is the magnitude-limited or flux-limited survey where many fainter objects are not detected.  In an unsupervised survey, this leads to `truncation' in the flux variable, such that nothing (not even the number) is known about the undetected population.  In a supervised survey, the astronomer seeks to measure a new property of a previously defined sample of objects.  Here nondetections produce `left-censored' data points: all of the objects are counted, but some have upper limits in the newly measured property.    The statistical treatment of censoring is well-established under the rubric of `survival analysis' as the problem arises (usually in the form of right-censoring) in fields such as biomedical research, actuarial science, and industrial reliability (Feigelson \& Nelson 1985).  The statistical treatment of truncation is more difficult as less is known about the full population, but some relevant methodology has been developed for astronomy.  

The `survival function' $S(x)$ for a univariate dataset is defined to be the inverse of the e.d.f.,
\begin{equation}
S(x) = P(X>x) =  \frac{\# {\rm observations} \ge x}{n} = 1 - F(x). 
\end{equation}
A foundation of survival analysis was the derivation of the nonparametric maximum likelihood estimator for a randomly censored dataset by Kaplan and Meier in the 1950s,
\begin{equation}
\hat{S}_{KM}(x) = \prod_{x_i \ge x} \left( 1 - \frac{d_i}{N_i} \right)
\end{equation}
where $N_i$ is the number of objects (detected or undetected) $\ge x_i$ and $d_i$ are the number of objects at value $x_i$.  If no ties are present, $d_i=1$ for all $i$. The ratio $d_i/N_i$ is the conditional probability that an object with value above $x$ will occur at $x$.  This `product-limit estimator' has discontinuous jumps at the detected values, but the size of the jumps increases at lower values of the variable because the weight of the nondetections are redistributed among the lower detections.  For large samples, the `Kaplan-Meier (KM) estimator' is asymptotically normal with variance
\begin{equation}
\widehat{Var}(\hat{S}_{KM}) = \hat{S}_{KM}^2 \sum_{x_i \ge x} \frac{d_i}{N_i(N_i-d_i)}.
\end{equation}

This nonparametric estimator is valid only when the censoring pattern is not correlated with respect to the variable $x$.  Note that when the estimator is used to obtain a luminosity function of a censored dataset, $S(L)$ and the censoring occurs due to a flux limit $f_0 = (L/(4 \pi d^2)^{1/2}$, the censoring pattern is only partially randomized depending on the distribution of distances in the sample under study.   There is no general formulation of an optimal nonparametric luminosity function in the presence of non-random censoring patterns.  However, if the parametric form of the luminosity function is known in advance, then estimation using maximum likelihood or Bayesian inference is feasible to obtain best-fit parameter values.  

It is possible to compare two censored samples with arbitrary censoring patterns without estimating their underlying distributions.  Several nonparametric hypothesis tests evaluating the null hypothesis $H_0 : S_1(x) = S_2(x)$ are available.  These include the Gehan and Peto-Peto tests (generalizations of the Wilcoxon two-sample test for censored data), the logrank test, and weighted Fleming-Harrington tests.  They all give mathematically correct probabilities that the two samples are drawn from the same distributions under different, reasonable treatments of the nondetections.  

A truly multivariate survival analysis that permits censoring in all variables has not been developed, but a few limited methods are available including generalizations of Kendall's $\tau$ rank correlation coefficient and various bivariate linear regression models.  Cox regression, which relates a single censored response variable to a vector of uncensored covariates, is very commonly used in biometrical studies.  

Truncation is ubiquitous in astronomical surveys as, except for a very few complete volume-limited samples, only a small portion of huge populations are available.  Unlike controlled studies in social sciences, where carefully randomized and stratified subsamples can be selected for measurement, the astronomer can identify only the closest and/or brightest members of a celestial population.  

As in survival analysis treating censored data, if the parametric form of the underlying distribution of a truncated dataset is known, then likelihood analysis can proceed to estimate the parameters of the distribution.  The nonparametric estimator equivalent to the Kaplan-Meier estimator was formulated by astrophysicist Donald Lynden-Bell in 1971, and studied mathematically by statistician Michael Woodroofe and others.  The Lynden-Bell-Woodroofe (LBW) estimator for a truncated dataset is similar to the Kaplan-Meier estimator, again requiring that the truncation values be independent of the true values.  A generalized Kendall's $\tau$ statistic is available to test this assumption.  A few extensions to the LBW estimator have been developed including a nonparametric rank test for independence between a truncated variable and a covariate, and a two-step least-squares regression procedure.  

The KM and LBW estimators, as nonparametric maximum likelihood estimators of censored and truncated distributions, are powerful tools.  Their performance is superior to heuristic measures often used by astronomers, such as the detection fraction or Schmidt's $1/V_{max}$ statistic.  However, care must be used: the dataset can not have multiple populations, the upper limit pattern may bias the result, results are unreliable when the censoring or truncation fraction is very high, and bootstrap resampling is often needed for reliable error analysis.  Furthermore, nonparametric survival methods cannot simultaneously treat nondetections and measurement errors in the detected points, which are linked by the instrumental process.  An integrated approach to measurement errors and nondetections is possible only within a parametric framework, as developed by Kelly (2007).

\subsection{Time series analysis}

The skys is filled with variable objects: orbits and rotations of stars and planets, stochastic accretion processes, explosive flares, supernovae, and gamma-ray bursts.  Some behaviors are strictly periodic, others are quasi-periodic, autocorrelated (including $1/f$-type red noise), or unique events.   Gravitational wave observatories are predicted to reveal all of these types of variations from high-energy phenomena, but to date only noise terms have been found.  Wide-field multi-epoch optical surveys of the sky, culminating with the planned LSST, make movies of the sky and will emerge with hundreds of millions of variable objects. 

Time series analysis is a vast, well-established branch of applied mathematics developed mostly in the fields of statistics, econometrics and engineering signal processing (Chatfield 2004, Shumway \& Stoffer 2006).  Astronomers have contributed to the methodology to address problems that rarely occur in these other fields: unevenly spaced observations, heteroscedastic measurement errors, and non-Gaussian noise.  Note that the mathematics applies to any random variable $x$ that is a function of a fixed time-like variable $t$, this includes astronomical spectra (intensities as a function of fixed wavelengths) and images (intensities as a function of fixed location).  

Unless periodicities are expected, analysis usually begins by examining the time domain observations, $x(t_i)$ where $i = 1, 2, \ldots, n$.   If the scientific goals seek large-scale changes in the values of $x$, then parametric regression models of the form $X(t) = f(t) + \epsilon(t)$ can characterize the trend, where the noise term is commonly modeled as a standard normal, $\epsilon = N(0,\sigma^2)$.  If trends are present but are considered uninteresting, they can be removed by a differencing filter, such as $y(t_i) = x(t_i) - x_{i-1}$, or spline fits.  

Very often, autocorrelation is present due either to astrophysical or instrumental causes.  The `autocorrelation function' as a function of lag time $t_k$ is
\begin{equation}
ACF(t_k) = \frac{\sum_{i=1}^{n-k} [x(t_i)-\bar{x}][x(t_{i+k})-\bar{x}]}{\sum_{i=1}^n [x(t_i)-\bar{x}]^2}.
\end{equation}
The `partial autocorrelation function' is very useful, as its amplitude gives the autocorrelation at lag $k$ removing the correlations at shorter lags.   Normal `white' noise produces ACF values around zero, while positive values indicate correlated intensities at the specified $t_k$ lag times.  When autocorrelation is present, the number of independent measurements is less than the number of observations $n$.  One effect on the statistics is to increase the uncertainty of standard statistics, such as the variance of the mean which is now  
\begin{equation}
\widehat{Var}[\bar{x}(t)] = \frac{\sigma^2}{n} \left[ 1 + 2 \sum_{k=1}^{n-1} \left( 1 -\frac{k}{n} \right) ACF(k) \right].
\end{equation}

Autocorrelated time series are usually modeled with stochastic `autoregressive models'.  The simplest is the random walk, 
\begin{equation}
x(t_i) = x(t_{i-1}) + \epsilon_i
\end{equation}
which gives an autocorrelation function that slowly declines with $k$, $ACF=1/\sqrt{1+k/i}$.  The random walk is readily generalized to the linear autoregressive (AR) model with dependencies on $p$ past values.  If the time series has stochastic trends, then a moving average (MA) terms is introduced with dependencies on $q$ past noise values.  The combination autoregressive moving average, ARMA($p,q$),  model is
\begin{equation}
x(t_i) = \alpha_1 x(t_{i-1}) + \ldots + \alpha_p x(t_{i-p}) + \epsilon(t_i) + \beta_1 \epsilon(t_{i-1}) + \ldots + \beta_q \epsilon(t_{i-q})
\end{equation}
This has been generalized to permit stronger trends,  long-term autocorrelations, multivariate time series with lags, non-linear interactions, heteroscedastic variances, and more  (ARIMA, FARIMA, VARIMA, ARCH, GARCH, and related models).   Model parameters are usually obtained by maximum likelihood estimation with the Akaike Information Criterion (equation \ref{aic.eqn}) for model selection.  This broad class of statistical models have been found be effective in modeling a wide range of autocorrelated time series for both natural and human-generated phenomena.  

A few methods are available for autocorrelated time series that are observed with uneven spacing.  A correlation function can be defined that involves binning either in time or in lag: the `discrete correlation function' (DCF) in astronomy (Edelson \& Krolik 1988) and the `slot autocorrelation function' in physics.  Binning procedures and confidence tests are debated, one recommendation is to apply Fisher's $z$ transform to the DCF to obtain an approximately normally-distributed statistic.   The `structure function' is a generalization of the autocorrelation function originally developed to study stochastic processes that can also can be applied to unevenly spaced data (Simonetti et al. 1986).  Structure functions, and the related singular measures, can give considerable insight into the nature of a wide variety of stochastic time series.  

A major mode of time domain analysis that is rarely considered in astronomy is `state-space modeling'.  Here the time series is modeled as a two-stage parametric regression problem with an unobserved  state vector defining the temporal behavior, a state matrix describing changes in the system, an observation matrix linking the data to the underlying state, and noise terms.   The model can be as complex as needed to describe the data: deterministic trends, periodicities, stochastic autoregressive behaviors, heteroscedastic noise terms, break points, or other components.  The parameters of the model are obtained by maximum likelihood estimation, and are updated by an efficient algorithm known as the `Kalman filter'.   Kitagawa \& Gersch (1996) illustrate the potential of state space modeling for problems in astronomy and geology.    

Spectral analysis (also called harmonic or Fourier analysis) examines the time series transformed into frequency space.  Periodic signals that are distributed in time but concentrated in frequency is now clearly seen.  The Fourier transform can be mapped to the $ACF$ according to 
\begin{equation}
f(\omega) =   \frac{\sigma_x^2}{\pi} \left[ 1 + 2 \sum_{k=1}^\infty ACF(k) ~{\rm cos} (\omega k) \right]
\end{equation}
This shows that, for white noise with $ACF=0$, the power spectrum $f(\omega)$ is equal to the signal variance divided by $\pi$.  If a phase-coherent sinusoidal signal is superposed, the $ACF(k)=cos(\omega_0 k)/(2\sigma_x^2)$ is periodic and the power spectrum is infinite at $\omega_0$.   An autoregressive $AR(1)$ process, the spectral density is large at low frequencies and decline at high frequencies, but may be more complicated for higher-order ARMA processes.  

Fourier analysis is designed for a very limited problem: a stationary time series of evenly spaced observations, infinite duration, with sinusoidal periodic signals superposed on white noise.  Any deviation from these assumptions causes aliasing, spectral leakage, spectral harmonics and splitting, red noise and other problems.     For realistic datasets, the classical `Schuster periodogram' is a biased estimator of the underlying power spectrum, and a number of techniques are used to improve its performance.  Smoothing (either in the time or frequency domain) increases bias but reduces variance, common choices are the Daniell (boxcar), Tukey-Hanning or Parzen windows.  Tapering, or reducing the signal at the beginning and end of the time series, decreases bias but increases variance.  The cosine or Hanning taper is common, and multitaper analysis is often effective (Percival \& Walden 1993).  

For unevenly spaced data, the empirical power spectrum is a convolution of the underlying process and the sequence of observation times.  The `Lomb-Scargle periodogram' (Scargle 1982) is a generalization of the Schuster periodogram for unevenly spaced data.  It can be formulated as a modified Fourier analysis, a least-squares regression to sine waves, or a Bayesian solution assuming a Jefferys prior for the time series variance.  An alternative to the Lomb-Scargle periodogram, which may have better performance at high frequencies, is the Direct Quadratic Spectrum Estimator (Marquardt \& Acuff 1984).  Astronomers have also developed a suite of non-Fourier periodograms for periodicity searches in unevenly spaced data.  These are statistics of the time domain data folded modulo a range of frequencies.  They include Dworetsky's `minimum string length', Lafler-Kinman's truncated autocorrelation function, Stellingwerf's `phase dispersion minimization', Schwarzenberg-Czerny's ANOVA statistic, and Gregory-Loredo's Bayesian periodogram.  Specialized methods are now being developed for the specific problem of uncovering faint planetary transits in photometric time series of stars.  

Evaluation of the significance of spectral peaks, both for traditional spectral analysis and for unevenly spaced data, is difficult.  Analytic approximations are available based on gamma, chi-square or $F$ distributions, but these are applicable only under ideal conditions when the noise is purely Gaussian and no autocorrelation is present.  Monte Carlo permutations of the data, simulations with test signals, and other methods (e.g. Reegen 2007) are needed to estimate false alarm probabilities and other characteristics of real-data power spectra.  

\subsection{Spatial point processes}

Spatial point datasets are a type of multivariate datasets where some of the variables can be interpreted as spatial dimensions.  Examples include locations of Kuiper Belt Objects in the 2-dimensional sky, galaxies in a 3-dimensional redshift survey (where recessional velocity is mapped into radial distance), photons on a 4-dimensional X-ray image (two sky dimensions, energy and arrival time), and Galactic halo stars in 6-dimensional phase space.  

Simple `spatial point processes' are stationary (properties invariant under translation) and isotropic (properties invariant under rotation). A stationary Poisson point process generates random locations; this pattern is called `complete spatial randomness' (CSR).  More complex models can be developed using non-stationary processes.   Data points may have associated non-spatial `mark variables': fluxes in some band, masses, velocities, classifications, and so forth.   Extensive methodology for interpreting such spatial point processes has been developed for applications in geology, geography, ecology and related sciences (Fortin \& Dale 2005, Illian et al. 2008).  Some of this work has been independently pursued in astronomy, particularly in the context of galaxy clustering (Mart\'inez \& Saar 2002).  

Astronomers have long experience in analyzing the `two-point correlation function' for characterize global autocorrelation in a stationary and isotropic spatial point process.   Let $P_{12}$ be the joint probability that two objects lie in ?infinitesimal spheres? of volume (area in two dimensions) $dV_1dV_2$ around two points at locations ${\bf x_1}$ and ${\bf x_2}$.  This probability can be modeled as the sum of a CSR process with spatial density $\bar{\rho}$ and a correlated process depending only on the distance $d$,
\begin{equation}
dP_{12} = \bar{\rho}^2 [1 + \xi(d)] dV_1 dV_2.
\end{equation}
Several estimators for $\xi$ have been investigated, often involving the ratio of observed densities to those in CSR simulations that have the same survey selection effects as the observed dataset.  Edge effects are particularly important for small surveys.  The variance of the $\xi$ estimator is often estimated using a normal approximation, but this is biased due to the spatial autocorrelation.  An alternative approach based on the block bootstrap is proposed by Loh (2008).  

Statisticians and researchers in other fields use integrals of the two-point correlation function  to avoid arbitrary choices in binning.  `Ripleys $K$ function' measures the average number of objects within a circle around each point,
\begin{equation}
K(d) = \frac{1}{\hat{\lambda} n} \sum_{i=1}^n \# [S~{\rm in}~ C({\bf s}_i,d)]
\end{equation}
where $C$ denotes a circle of radius $s$ centered on the data point locations ${\bf x}_i$ and $\lambda$ measures the global density.  For a CSR process, $\lambda = n/V$ where $V$ is the total volume of the survey, $E[K(d)] = \pi d^2$ and $Var[K(d)[ = 2 \pi d^2 / (\widehat{\lambda^2} V)$.   To remove the rapid rise as circle radii increase, the stabilized function $L = \sqrt{K(d)/\pi}-d$  (for two dimensions) with uniform variance is often considered.  These statistics can easily be extended to measure clustering interactions between two or more populations.

As with the two-point correlation function, the $K$ function is biased due to edge effects as the circle radii become comparable to the survey extent.  Various edge-corrections have been developed.  Other related functions have different sensitivities to clustering and edge effects.  The $G(d)$ function is the cumulative distribution of nearest neighbor distances computed from the data points, and $F(d)$ is the similar distribution with respect to random locations in the space.  $F$ is called the `empty space function' and serves the same role as the CSR simulations in astronomers' estimates of $\xi$.  `Baddeley's $J$ function', defined by $J(d) = [1-G(d)] / [1-F(r)]$, is highly resistant to edge effects while still sensitive to clustering structure: $J(d)=1$ for CSR patterns, $J(r)<1$ for clustered patterns out to the clustering scale, and $J(r)>1$ for patterns with spatial repulsion (e.g., a lattice).  Figure~\ref{PSSS_Feigelson_spatial.fig} shows an application of Baddeley's $J$ to a galaxy spatial distribution.  

All of the above statistics measure global patterns over the entire space, and thus are applicable only to stationary spatial processes.  In a nonstationary point process, the clustering pattern varies across the observed space.  Moran's $I$ can be applied locally to the $k$-nearest neighbors, giving an exploratory mapping tool for spatially variable autocorrelation.  Probabilities for statistical testing are now difficult to obtain, but block bootstrap methods that maintain small-scale structure but erase large-scale structure might be useful.

In some situations, the spatial variables are themselves less interesting than the mark variables which represent an unseen continuous process that has been sampled at distinct spatial locations.   For example, individual background galaxies distorted by gravitational lensing trace an unseen foreground Dark Matter distribution, and individual star radial velocities trace a global differential rotation in the Galactic disk.   An important method for interpolating mark variables was developed in geology under the rubric of `kriging'.  For a stationary Gaussian spatial point process, kriging gives the minimum square error predictor for the unobserved continuous distribution.  Kriging estimates are based on semi-variogram autocorrelation measures, usually using maximum likelihood methods.

\section{Resources}

The level of education in methodology among astronomers (and most physical scientists) in available statistics is far below their needs.  Three classes of resources are needed to advance the applications of statistics for astronomical research:  education in existing methodology, research in forefront astrostatistics, and advanced statistical software.  In the U.S., most astronomy students take no classes in statistics while the situation is somewhat better in some other countries.  Brief summer schools exposing students to methodology are becoming popular, and partially alleviate the education gap. 

Self-education in statistics is quite feasible as many texts are available and astronomers have the requisite mathematical background for intermediate-level texts.  The Wikipedia Web site is quite effective in presenting material on statistical topics in a compact format.   But astronomers confronting a class of statistical problem would greatly benefit from reading appropriate textbooks and monographs.   Recommended volumes are listed in Table~\ref{statbooks.tbl}. 

Research in astrostatistics has rapidly increased since the mid-1990s.  Dozens of studies to treat problems in unevenly spaced time series, survey truncation, faint source detection, fluctuations in the cosmic background radiation, and other issues have been published in the astronomical literature.  Some of these efforts develop important new capabilities for astronomical data analysis.  However, many papers are not conversant with the literature in statistics and other applied fields, even at the level of standard textbooks.   Redundant or inadequate treatments are not uncommon.  

The paucity of software within the astronomical community has, until recently, been a serious hurdle to the implementation of advanced statistical methodology.  For decades, advanced methods were provided only by proprietary statistical software packages like {\it SAS} and {\it S-Plus} which the astronomical community did not purchase.  The commonly purchased {\it IDL} package for data analysis and visualization incorporates some statistical methods, particularly those in {\it Numerical Recipes} (Press et al. 1986), but not a full scope of modern methods.  

The software situation has enormously improved in the past few years with the growth of {\bf R}, an implementation of the {\it S} language released under GNU Public License.  {\bf R} is similar to {\it IDL} in style and operation, with brief user commands to implement both simple and complex functions and graphics.   Base {\bf R} implements some dozens of standard statistical methods, while its user-supplied add-on packages in the Comprehensive {\bf R} Archive Network ({\bf CRAN}) provide thousands of additional functionalities.  Together, {\bf R/CRAN} is a superb new resource for promulgation of advanced statistical methods into the astronomical community.  Its use is illustrated below, and list a few of the dozens of books recently emerged on use of {\bf R} for various purposes.  

\subsection{Web sites and books}

The field of statistics is too vast to be summarized in any single volume or Web site. {\it Wikipedia} covers many topics, often in an authoritative fashion (see \\ http://en.wikipedia.org/wiki/List\_of\_statistics\_articles).   Additional on-line resources specifically oriented towards astrostatistics are provided by Penn State's Center for Astrostatistics (http://astrostatistics.psu.edu), the California-Harvard Astrostatistics Collaboration (http://www.ics.uci.edu/$\sim$dvd/astrostat.html), and the International Computational AstroStatistics Group (http://www.incagroup.org).  The International Statistics Institute has formed a new Astrostatistics Network to foster cross-disciplinary interaction (http://isi-web.org/com/ast).  

The greatest resource are the hundreds of volumes written be statisticians and application experts.  Table~\ref{statbooks.tbl} gives a selection of these books recommended for astronomers.  They include undergraduate to graduate level texts, authoritative monographs, and application guides for the {\bf R} software system.

\newpage
\centerline{\large\bf Selected statistics books for astronomers}
\medskip

{\small
\begin{tabular}{| ll |} \hline
\label{statbooks.tbl}
& \\
\multicolumn{2}{|  l |}{\bf Broad scope} \\
Adler (2018) & R in a Nutshell \\
Dalgaard (2008) & Introductory Statistics with R \\
Feigelson \& Babu (2011) & Modern Statistical Methods for Astronomy with R \\
     & Applications \\
Rice (1994) & Mathematical Statistics and Data Analysis \\
Wasserman (2004) & All of Statistics: A Concise Course in \\
    & Statistical Inference \\
& \\
\multicolumn{2}{|  l |}{\bf Statistical inference} \\
Conover (1999) & Practical Nonparametric Statistics \\
Evans et al. (2000) & Statistical Distributions \\
Hogg \& Tanis (2009) & Probability and Statistical Inference \\
James (2006) & Statistical Methods in Experimental Physics \\
Lupton (1993) & Statistics in Theory and Practice \\
Ross (2009) & A First Course in Probability \\
& \\
\multicolumn{2}{|  l |}{\bf Bayesian statistics} \\
Gelman et al. (2003) & Bayesian Data Analysis \\
Gregory (2005) & Bayesian Logical Data Analysis for the Physical Sciences \\
Kruschke (2011) & Doing Bayesian Data Analysis: A Tutorial with R and BUGS \\
& \\
\multicolumn{2}{|  l |}{\bf Resampling methods} \\
Efron \& Tibshirani (1993) & An Introduction to the Bootstrap \\
Zoubir \& Iskander (2004) & Bootstrap Techniques for Signal Processing \\
& \\
\multicolumn{2}{|  l |}{\bf Density estimation} \\
Bowman \& Azzalini (1997)& Applied Smoothing Techniques for Data Analysis \\
Silverman (1998) & Density Estimation \\
Takezawa (2005) & Introduction to Nonparametric Regression \\
& \\
\multicolumn{2}{|  l |}{\bf Regression and multivariate analysis}\\
Kutner et al. (2004) & Applied Linear Regression Models \\
Johnson \& Wichern (2007) &  Applied Multivariate Statistical Analysis \\
\hline
\end{tabular}

\begin{tabular}{| ll | } \hline

\multicolumn{2}{|  l |}{\bf Nondetections} \\
Helsel (2005) & Nondetects and Data Analysis \\
Klein \& Moeschberger (2010) & Survival Analysis \\ 
Lawless (2002) & Statistical Models and Methods for Lifetime Data \\
& \\
\multicolumn{2}{|  l |}{\bf Spatial processes} \\
Bivand et al. (2008) & Applied Spatial Data Analysis with R \\
Fortin \& Dale (2005) & Spatial Analysis: A Guide for Ecologists \\
Illian et al. (2008) & Statistical Analysis and Modelling of Spatial \\ 
     & Point Patterns \\
Mart\'inez \& Saar (2002) & Statistics of the Galaxy Distribution \\
Starck \& Murtagh (2006) & Astronomical Image and Data Analysis \\
& \\
\multicolumn{2}{|  l |}{\bf Data mining, clustering, and classification} \\
Everitt et al. (2001) &  Cluster Analysis \\
Duda et al. (2001) & Pattern Classification \\
Hastie et al. (2001) & The Elements of Statistical Learning \\ 
& \\
\multicolumn{2}{|  l |}{\bf Time series analysis} \\
Chatfield (2004) & The Analysis of Time Series \\
Cowpertwait \& Metcalfe (2009) & Introductory Time Series with R \\
Nason (2008) & Wavelet Methods in Statistics with R \\
Shumway \& Stoffer (2006) & Time Series Analysis and Its Applications with\\
    & R Examples \\
 & \\
\multicolumn{2}{|  l |}{\bf Graphics and data visualization} \\
Chen et al. (2008) & Handbook of Data Visualization \\
Maindonald (2010) & Data Analysis and Graphics using R \\
Sarkar (2008) & Lattice: Multivariate Data Visualization with R \\
Wickham (2009) & ggplot2: Elegant Graphics for Data Analysis \\
\hline
\end{tabular}
 \newpage
}

\subsection{The {\bf R} statistical software system}

{\bf R} (R Development Core Team, 2010) is a high-level software language in the public domain with {\bf C}-like syntax.  It provides broad capabilities for general data manipulation with extensive graphics, but its strength are the dozens of built-in statistical functionalities.    Compiled code for {\bf R} can be downloaded for Windows, MacOS and Unix operating systems from http://r-project.org. The user community of {\bf R} is estimated to be 1-2 million individuals, and over 70 books have been published to guide researchers in its use.

The {\bf Comprehensive R Archive Network} ({\bf CRAN}) of user-supplied add-on packages  has been growing nearly-exponentially for a decade, currently with $\sim 3000$ packages. Some {\bf CRAN} packages have narrow scope, while others are themselves large statistical analysis systems for specialized analyses in biology, ecology, econometrics, geography, engineering, and other research communities.  {\bf CRAN} packages, as well as user-provided code in other languages (C, C++, Fortran, Python, Ruby, Perl, Bugs, and XLisp),  can be dynamically brought into an {\bf R} session. Some {\bf CRAN} packages can be productively used in astronomical science analysis.  At the time of writing, only one package has been specifically written for astronomy: {\bf CRAN}'s {\it fitsio} package for input of  FITS (Flexible Image Transport System, http://fits.gsfc.nasa.gov/) formatted data used throughout the astronomical community.  

The {\bf R} scripts below, and accompanying figures, illustrate the analysis and graphical output from {\bf R} and {\bf CRAN} programs based on an astronomical dataset. Similar {\bf R} scripts for a wide variety of statistical procedures applied to astronomical datasets are given in Feigelson \& Babu (2011).  

The first {\bf R} script inputs a multivariate dataset of photometric and other measurements in ASCII using {\bf R}'s {\it read.table} function. The data are extracted from the Sloan Digital Sky Survey quasar survey (Schneider et al. 2010).  The {\it dim} (dimension) function tells us there are 17 columns for 33,845 quasars,  {\it names} give the column headings, and {\it summary} gives the  minimum, quartiles, mean and maximum value for each variable.  The next lines select the first 200 quasars and define redshift and $r-i$ color index variables.  

The second script makes  a boxplot  summarizing the redshift distribution.  It gives the median with a horizontal line, 25\% and 75\% quartiles with `hinges', `whiskers', outliers, and notches representing a nonparametric version of the standard deviation.  It shows an asymmetrical distribution with a heavy tail towards higher redshifts around $z \simeq 3-4$.  This is one of the common plots made within {\bf R}.  

\begin{figure}
\centering
\includegraphics[height=01.0\textwidth,angle=-90.]{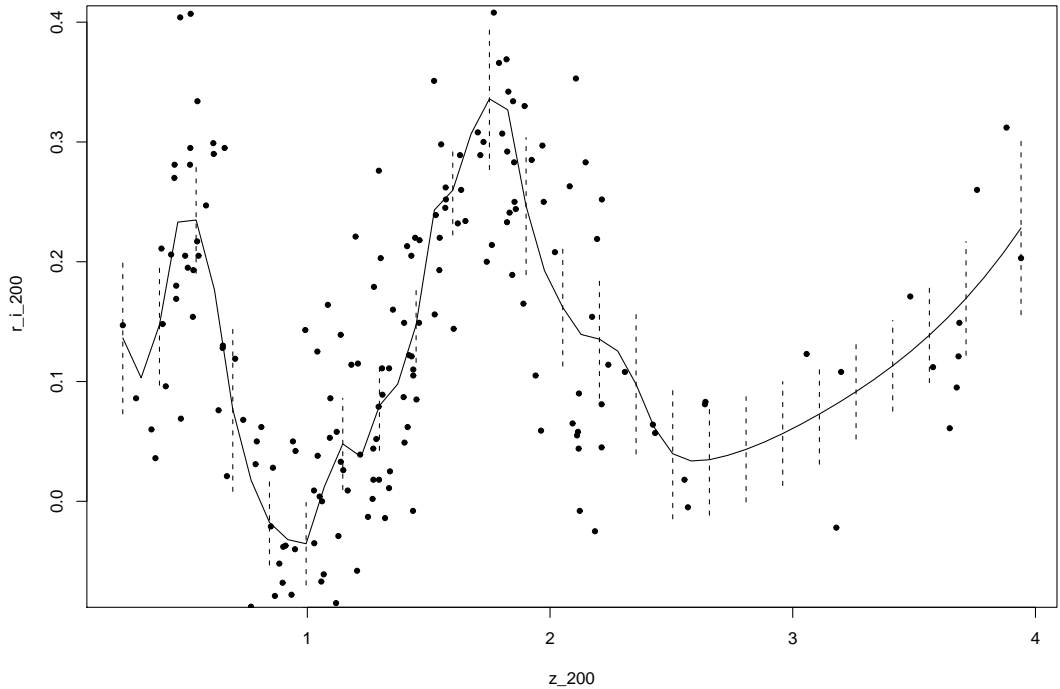}
\caption{Adaptive kernel density estimator of quasar $r-i$ colors as a function of redshift with bootstrap confidence intervals, derived using the {\bf np} package, one of $\sim 3000$ add-on {\bf CRAN} packages.} 
\label{PSSS_Feigelson_adapt.fig}
\end{figure}

The third script calculates a sophisticated, computationally intensive, nonparametric kernel smoother to the two-dimensional distribution of the $r-i$ color index dependence on redshift $z$.  This calculation is provided by the {\bf CRAN} package {\it np}, {\it Nonparametric kernel methods for mixed datatypes} (Hayfield \& Racine 2008).  Its {\it npregbw} function computes an adaptive kernel density estimator for a $p$-dimensional matrix of continuous, discrete and/or categorical variables.  The options chosen here give a local-linear regression estimator with a $k$-nearest neighbor adaptive bandwidth obtained using least-squares cross-validation.  Figure~\ref{PSSS_Feigelson_adapt.fig} shows the quasar color-redshift data points superposed on the kernel smoother with error bars representing the 95\% confidence band based on bootstrap resampling.  The methods are outlined in the {\bf CRAN} help file and are described in detail by Hall et al. (2007). 

\begin{verbatim}

# 1. Construct a sample of 200 SDSS quasar redshifts and r-i colors 

qso <- read.table('http://astrostatistics.psu.edu/datasets/SDSS_QSO.dat', 
    header=T) 
dim(qso) 
names(qso) 
summary(qso) 
z_200 <- qso[1:200,4]  
r_i_200 <- qso[1:200,9] - qso[1:200,11] 

# 2. Univariate distribution of redshifts: boxplot 

boxplot(z_200, varwidth=T, notch=T, main='SDSS quasars', ylab='Redshift', 
   pars=list (boxwex=0.3,boxlwd=1.5,whisklwd=1.5,staplelwd=1.5,outlwd=1.5,font=2))

# 3. Bivariate adaptive kernel estimator with bootstrap errors
 
install.packages('np') 
library(np) 
citation('np')
bw_adap <- npregbw(z_200, r_i_200, regtype='ll', bwtype='adaptive_nn')
npplot(bw_adap, plot.errors.method="bootstrap")
points(z_200, r_i_200, pch=20)
\end{verbatim}

To illustrate an analysis of galaxy clustering, the fourth {\bf R} script inputs 4,215 galaxies from a redshift survey of the Shapley supercluster by Drinkwater et al. (2004).  A subsample of 286 galaxies is selected from the original sample by specifying a small location in the sky.  

\begin{figure}
\centering
\vspace{-1in}
\includegraphics[angle=-90.,width=1.0\textwidth]{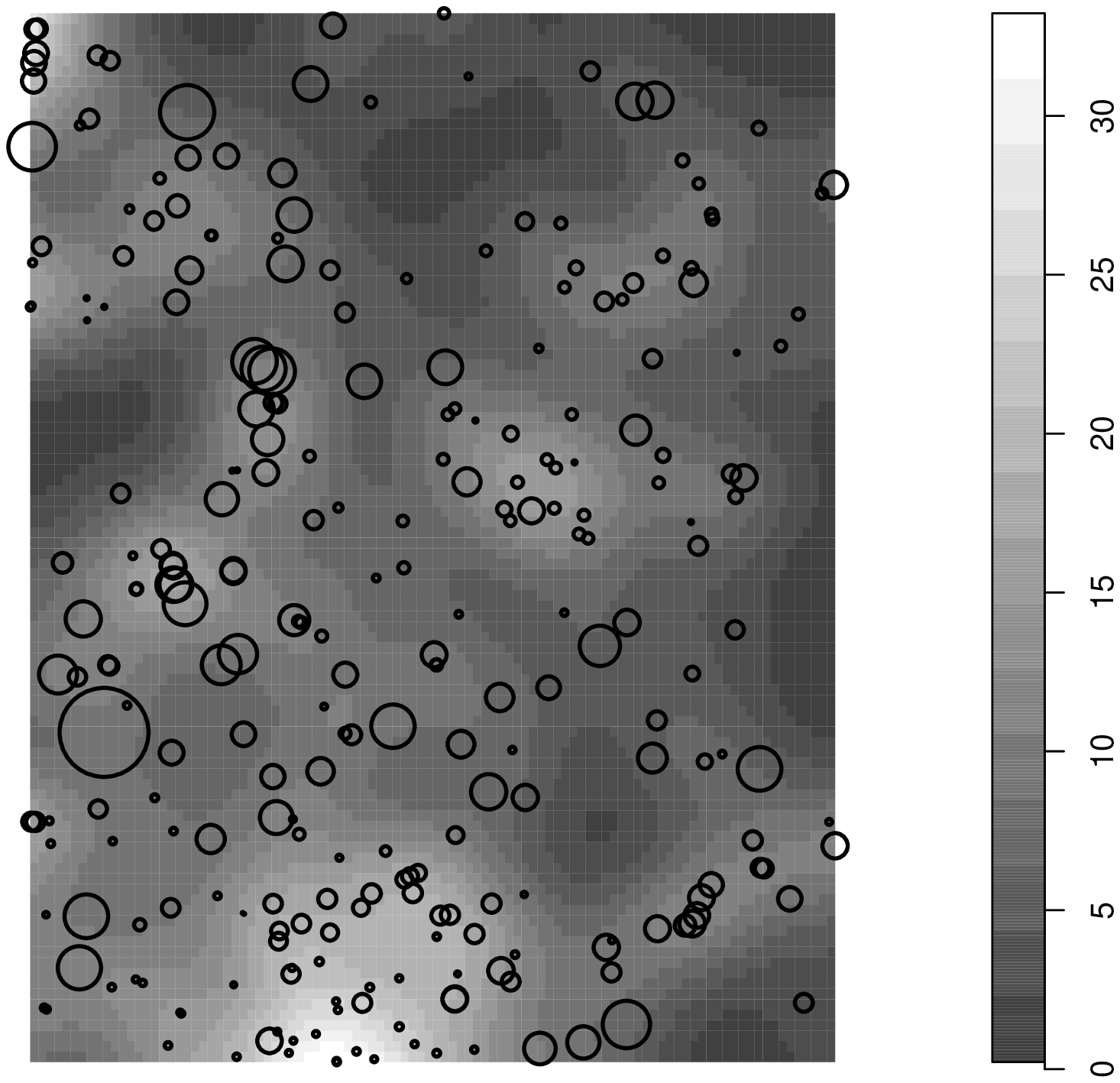} \\
\vspace{-0.5in}
\includegraphics[angle=-90.,width=0.5\textwidth]{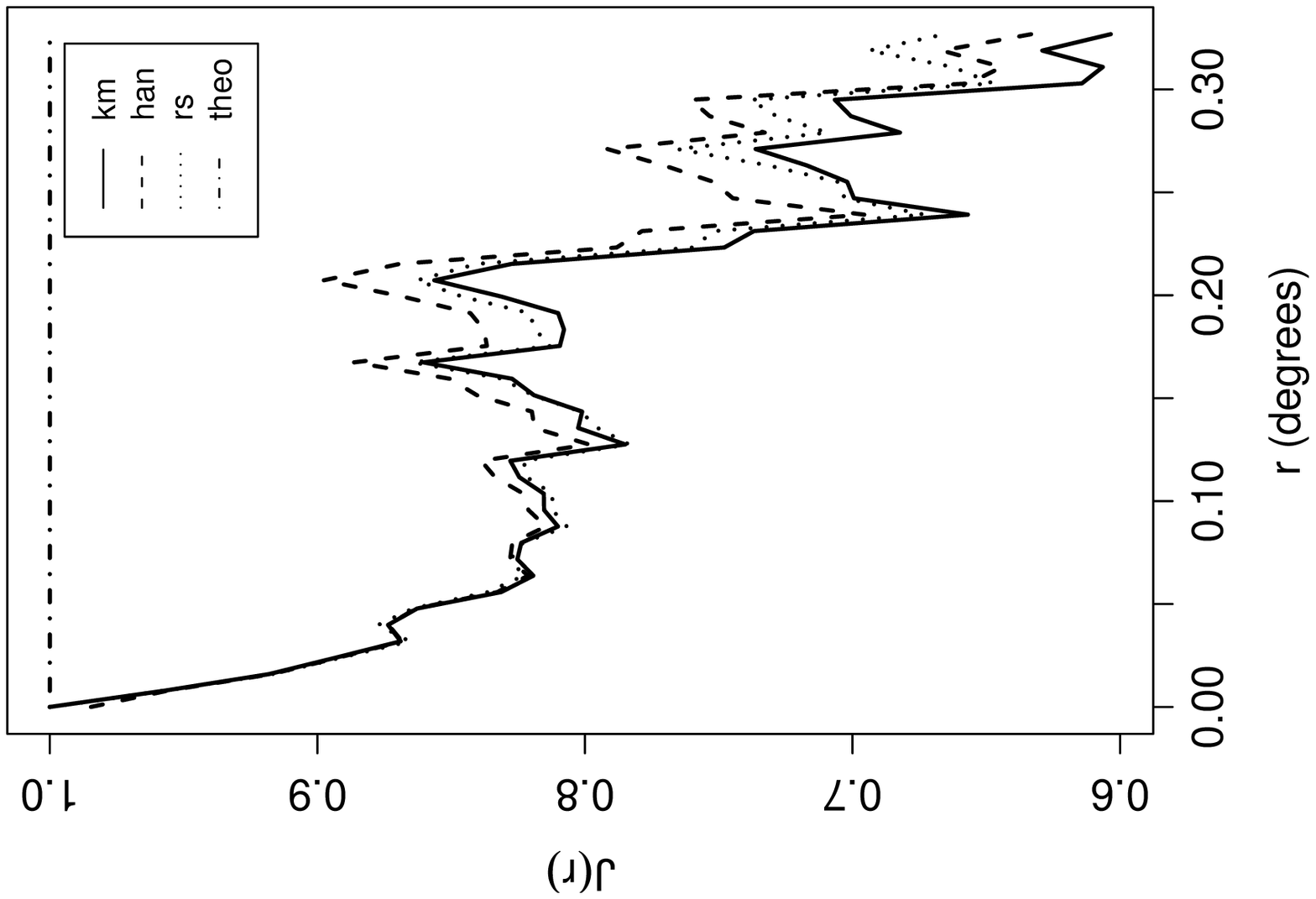}
\caption{Analysis of a small galaxy redshift survey as a spatial point process using {\bf CRAN}'s {\it spatstat} package.  Left: Sky locations of the galaxies with circle sizes scaled to their redshifts, superposed on a kernel density estimator of the sky positions.  Right: Baddeley's $J$ function derived from this galaxy sample.  $J$ is related to the integral of the two-point correlation function, and is shown with three edge corrections. } \label{PSSS_Feigelson_spatial.fig}
\end{figure}

The fifth {\bf R} script starts by installing the {\bf CRAN} {\it spatstat}, {\it Spatial Statistics}, package described by Baddeley \& Turner (2005).   The functions {\it owin} and {\it as.ppp} convert the data table into a special format used by {\it spatstat}.  The galaxies are then plotted on a smoothed spatial distribution with symbol sizes scales to the recessional velocity as a `mark' variable.  The second-order $J$ function, calculated by the function $Jest$, is then plotted showing sensitivity at large angles to different edge correction algorithms (Figure~\ref{PSSS_Feigelson_spatial.fig}).  The horizontal line is the predictor for `complete spatial randomness'.  Baddeley's $J$ function is related to Ripley's $K$ function, the cumulative (unbinned) two-point correlation function, and is designed to reduce edge effects.

\begin{verbatim}

# 4. Input and examine Shapley galaxy dataset

shap <- read.table('http://astrostatistics.psu.edu/datasets/Shapley_galaxy.dat',
   header=T)
attach(shap) 
dim(shap) 
summary(shap)
shap_lo <- shap[(R.A.<214) & (R.A.>209) & (Dec.>-34) & (Dec.<-27),]
dim(shap_lo)

# 5. Display galaxy distribution and calculate Baddeley J function 
# using spatstat package

install.packages('spatstat') 
library(spatstat)
citation(spatstat)
shap_lo_win <- owin(range(shap_lo[,1]), range(shap_lo[,2]))
shap_lo_ppp <- as.ppp(shap_lo[,c(1,2,4)], shap_lo_win) 
summary(shap_lo_ppp)

par(mfrow=c(1,2))
plot(density(shap_lo_ppp, 0.3), col=gray(5:20/20), main='')
plot(shap_lo_ppp, lwd=2, add=T)
plot(Jest(shap_lo_ppp), lwd=2, col='black', cex.lab=1.3, cex.axis=1.3,
  main='',xlab='r (degrees)', legendpos='topright')
par(mfrow=c(1,1))
\end{verbatim}

~\\
ACKNOWLEDGEMENTS \\
The work of Penn State's Center for Astrostatistics is supported by NSF grant SSE AST-1047586.

\end{document}